\documentclass[preprint,11pt]{revtex4-1}
\usepackage[utf8]{inputenc}
\usepackage[T1]{fontenc}
\usepackage{bm}
\usepackage{amsmath}
\usepackage{amsfonts}
\usepackage{amssymb}
\usepackage{graphicx}
\usepackage{float}
\makeatletter
\let\saved@includegraphics\includegraphics
\AtBeginDocument{\let\includegraphics\saved@includegraphics}
\makeatother
\RequirePackage{siunitx}
\usepackage{xcolor}
\newcommand*{\dif}{\,\mathrm{d}}
\newcommand{\ms}{\si{\micro\metre\per\second}}
\newcommand{\mum}{\si{\micro\metre}}
\begin{document}
\title{Spontaneous vortex formation by microswimmers with retarded attractions}
\author{Xiangzun Wang$^{1}$, Pin-Chuan Chen$^{2}$, Klaus Kroy$^{2}$, Viktor Holubec$^{3}$ and Frank Cichos${}^{1\ast}$\\
\textit{${}^{1}$Peter Debye Institute for Soft Matter Physics, Leipzig University, 04103 Leipzig, Germany.}\\
\textit{${}^{2}$Institute for Theoretical Physics, Leipzig University, Postfach 100 920, 04009 Leipzig, Germany,}\\
\textit{${}^{3}$Department of Macromolecular Physics, Faculty of Mathematics and Physics, Charles University, 18000~Prague, Czech Republic,}\\
\textit{$^\ast$To whom correspondence should be addressed; E-mail:  cichos@physik.uni-leipzig.de.}
}

\begin{abstract}
Collective states of inanimate particles self-assemble through physical interactions and thermal motion. Despite some phenomenological resemblance, including signatures of criticality, the autonomous dynamics that binds motile agents into flocks, herds, or swarms allows for much richer behavior. Low-dimensional models have hinted at the crucial role played in this respect by perceived information, decision-making, and feedback implying that the corresponding interactions are inevitably retarded.
Here we present experiments on spherical Brownian microswimmers with delayed self-propulsion toward a spatially fixed target. We observe a spontaneous symmetry breaking to a transiently chiral dynamical state and concomitant critical behavior  that does not rely on many-particle cooperativity. By comparison with the stochastic delay differential equation of motion of a single swimmer, we pinpoint the delay-induced effective synchronization of the swimmers with their own past as its key mechanism. Increasing numbers of swimmers self-organize into layers with pro- and retrograde orbital motion, synchronized and stabilized by steric and hydrodynamic interactions. Our results demonstrate how even most simple retarded non-reciprocal interactions can foster emergent complex adaptive behavior in relatively small ensembles.
\end{abstract}

\maketitle
\section*{Introduction}
Ordered dynamical phases of motile organisms are ubiquitous in nature across all scales \cite{Kauffman.1993}, from bacterial colonies, to insect swarms and bird flocks \cite{Vicsek.2012}. In particular, self-organization into vortex patterns is often observed and has been attributed to some local external attractor, e.g., light or nutrient concentration, together with behavioral rules like collision avoidance and mutual alignment \cite{Delcourt.2016}.  
Pertinent social interactions are commonly thought to be based on perception \cite{Strandburg-Peshkin.2013, Pearce.2014,Cremer.2019} and the ability to actively control the direction of motion \cite{Delcourt.2016}. They are also generally presumed to provide some benefits to the individual and to the collective, as in the case of collision avoidance or predator evasion \cite{Couzin.2003,Ioannou.2012}. However, since such interactions are usually derived only indirectly and approximately from observations \cite{Berdahl.2018i5}, it is arguably useful to coarse grain them, e.g., into simple alignment rules.  This strategy has been successful in physics in order to rationalize complex collective effects with the help of simple mechanistic models, in particular with respect to emerging universal traits \cite{Vicsek.1995,Hemelrijk.2011,Costanzo.2018,Delcourt.2016}. It is also supported by the observation that biological many-body assemblies often appear highly susceptible to environmental influences and exhibit a dynamical finite-size scaling reminiscent of critical states in inanimate physical systems \cite{Cavagna.2010,Mora.2011,Munoz.2018,Cavagna.2017ma}. 

Importantly, the cascades of complex biochemical/biophysical processes \cite{Kim,Zhang.2019pcp} needed to transform signal perception into a navigational reaction inevitably result in retarded interactions upon coarse-graining~\cite{More.2018}. This generic complication is often dismissed in the analysis, and dedicated models and experiments addressing the role of time delays in active matter are still rare \cite{Mijalkov.2016,Forgoston.2008,Piwowarczyk.2019,Khadka.2018}, although these have occasionally been shown to fundamentally alter the collective dynamics \cite{Forgoston.2008} and to bring it closer to that found in nature \cite{Holubec.2021}. To a first approximation, delay effects can resemble inertial corrections to an otherwise overdamped biological dynamics \cite{Attanasi.2014}. In particular, both have a propensity to give rise to oscillations, and inertia moreover to rotational motion around an attractive center, as familiar from planetary orbits. 

Experiments that can assess or even deliberately control retarded interactions in living systems turn out to be difficult. But by imposing time delays onto synthetic active particles via computer control, we can create an ideal laboratory system to experimentally emulate such situations. Suitable feedback control techniques for active particles have recently become available through photon nudging  \cite{Qian.2013zq}. The technique allows to adjust a particle's propulsion speed to acquired real-time information (positions, directions of motion) about the dynamical state of an ensemble. It has previously been employed to rectify the rotational Brownian motion for particle steering and trapping \cite{Bregulla.2014}, to explore  orientation-density patterns in activity landscapes \cite{Soeker.2021}, and to study information flow between active particles \cite{Khadka.2018} and their emerging critical states \cite{Baeuerle.20204d8,Loeffler.2021}. Beyond what related computer simulations accomplish \cite{Liebchen.2019h47,Stark.2018,Auschra.2021pnk}, these experiments additionally incorporate the full real-world complexity arising from actual physical interactions due to hydrodynamic, thermal, or concentration fields.  
In the following we describe experiments with feedback-controlled active Brownian microswimmers ``aiming'' to reach a fixed target by a retarded thermophoretic self-propulsion. The resulting systematic navigational ``errors'' are seen to cause a spontaneous symmetry breaking to a bi-stable dynamical state, in which the swimmers self-organize into a merry-go-round motion that switches transiently between degenerate chiralities. 

\begin{figure}[!ht]
    \centering
    \includegraphics[width=1.0\columnwidth]{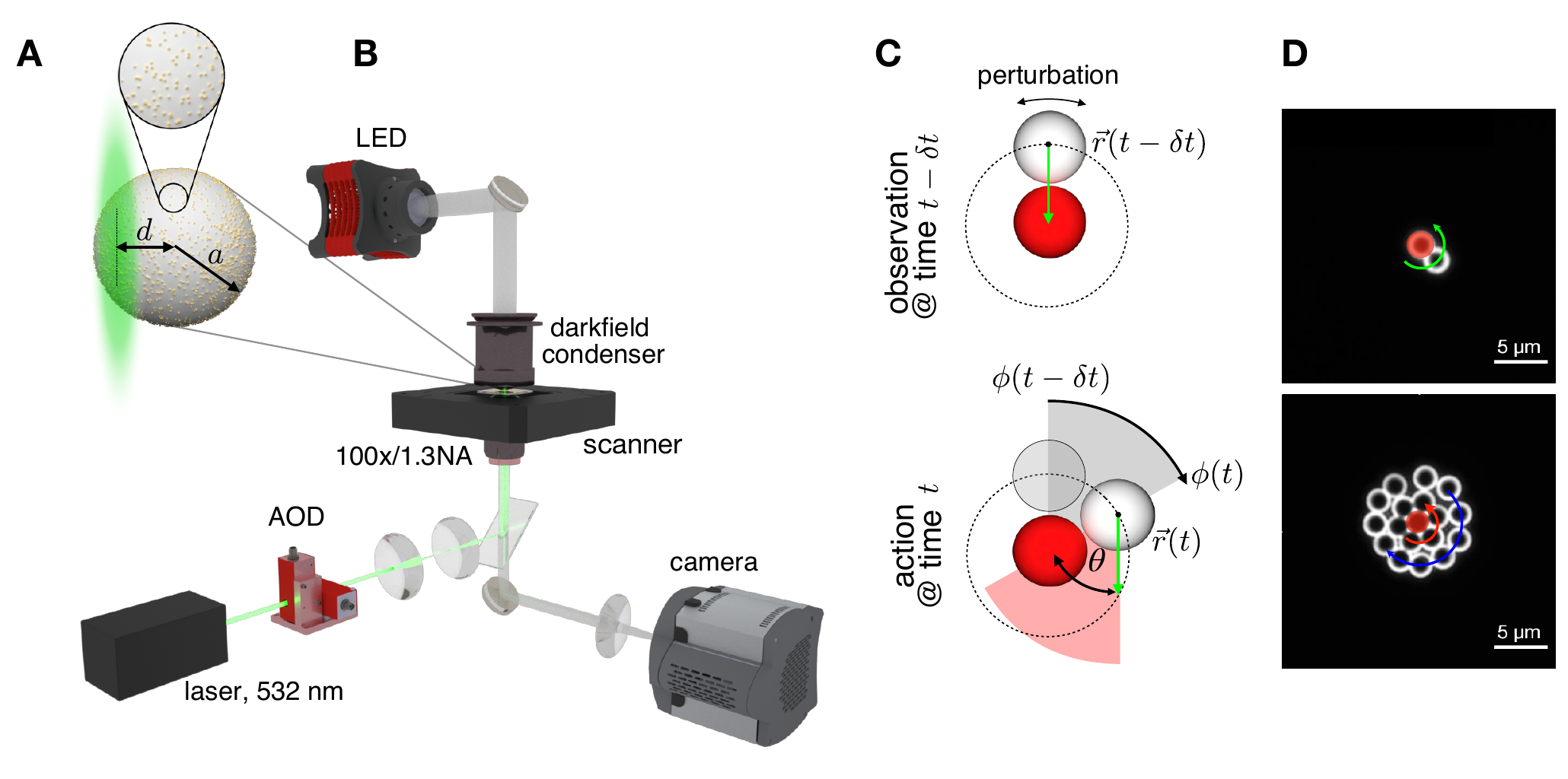}
    \caption{\small{\bf Experimental Realization:} {\bf A} Particles used in the experiments consist of a melamine resin colloid (2.18\,$\mum$ in diameter) with 8\,nm gold nanoparticles scattered across the surface (covers up to 10\% of the total surface area). A 532\,nm laser focused at the edge of the particle at a distance $d$ from its center induces a self-thermophoretic motion and allows for precise control of the propulsion direction. {\bf B}. Experimental setup used to control active particles. Active particles are imaged by darkfield microscopy (LED, darkfield condenser and camera). The particle control is achieved by the laser with a 2-axis acousto-optic deflector (AOD) that allows sequential beam steering of the laser on the sample plane. All particles in the field of view are addressed during each exposure period of the camera. {\bf C} The interaction rule for the delayed attraction of a single active particle (white sphere) towards a target (red sphere) is split into an observation made at a time $t-\delta t$ that sets the direction of motion for the self-propulsion step exerted after a programmed delay time $\delta t$. The green arrows indicate the direction $-\vec{r}_i(t-\delta t)$ of propulsion at time $t$ (the origin is at the center of the target). {\bf D} Examples of darkfield microscopy images where a single active particle (top) and 16 active particles (bottom) interact with one target particle (red). \label{fig:figure1}}
\end{figure}

\section*{Results}
\paragraph*{Single Particle Retarded Interaction}
The basic component of a swarm is a single active particle whose direction of motion depends dynamically on its environment. A perturbation of the particle position leads to an adjustment of the direction of motion, which can inevitably only take place with a delay between detection and reaction to the perturbation. In this work, we study the simplest case, where an active particle moves toward an immobilized target particle of the same size. Assuming that the active particle has a programmed time delay $\delta t$ in its response, its propulsion direction $\hat{\mathbf{u}}(t)$ at time $t$ is determined by its relative position to the target particle it "sensed" at time $t-\delta t$ in the past according to
\begin{equation}\label{eq:target_cohesion}
    \hat{\mathbf{u}}(t) = \frac{-\mathbf{r}(t-\delta t)}{|\mathbf{r}(t-\delta t)|},
\end{equation}
where $\mathbf{r}$ is the location of the active particle with respect to the target particle at the origin. 
This interaction rule is implemented in an experimental feedback system that controls the propulsion of active particles. Here, the active particles are polymer particles with a radius of $a=1.09\,\mum$ coated with gold nanoparticles and suspended in a thin film of water. A laser with a wavelength of 532\,nm is focused on the active particle with a displacement $d$ from the center (Fig.~\ref{fig:figure1}B) to drive the particle at a constant velocity $v_0$ in the direction defined by Eq.~\ref{eq:target_cohesion} via self-thermophoresis \cite{Fraenzl.2021}. A darkfield microscopy setup is used to image the particles (Fig.~\ref{fig:figure1}C). A feedback loop running on a computer analyzes and records the positions of the particles to control the laser position accordingly via an acousto-optical deflector. We use a separate calibrator particle running on a quadratic trajectory as a reference for the velocity $v_0$ of a free particle. Further details are described in the Methods section and the Supporting Materials.

\begin{figure*}[h!]
    \centering
    \includegraphics[width=0.8\columnwidth]{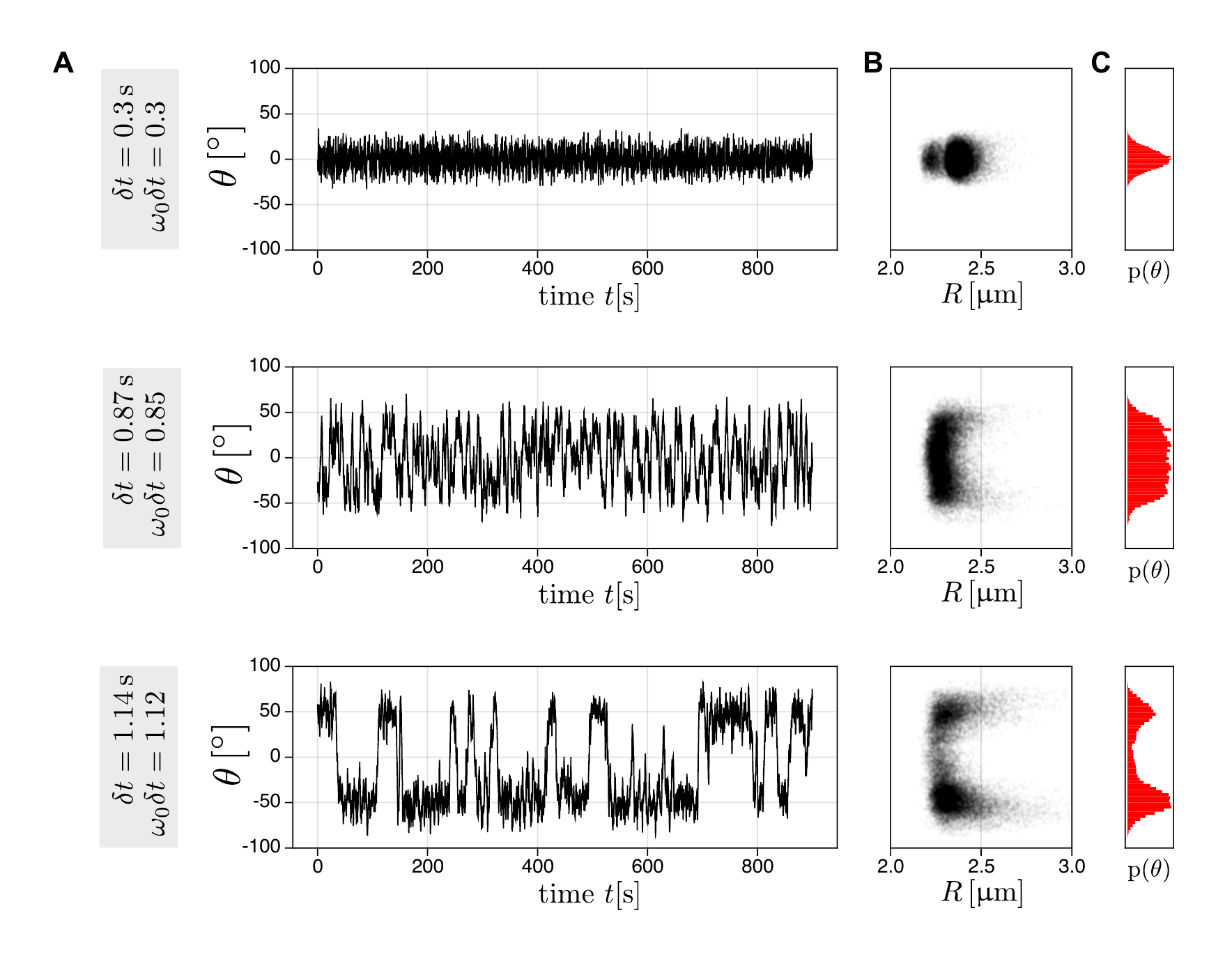}
    \caption{{\bf Propulsion angle at different programmed delay:} {\bf A} Trajectories of the propulsion angle $\theta(t)$ of an active particle at three different delays (top: $\delta t=0.3\, {\rm s}$, middle: $\delta t=0.87\, {\rm s}$ and bottom: $ \delta t=1.14 \, {\rm s})$  for its attraction towards a target particle. The velocity of the active particle is $v_0$ = 2.16 $\ms$. {\bf B:}
    Trajectory points of the propulsion angle $\theta(t)$ vs. the distance $R$ of the particle from the target particle.
    {\bf C:} Histograms of the propulsion angle over the whole trajectory. The delay for the individual panels in columns {\bf B} and {\bf C} is indicated on the left of the corresponding row.
    \label{fig:figure2} }
\end{figure*}

When the system is started with a programmed time delay $\delta t=0\,{\rm s}$, the active particle moves towards the target particle until it collides with it. The motion of the active particle is then constrained by the target particle carrying out a diffusive motion around the target's circumference. Its distance to the target particle obeys the barometric distribution \cite{Selmke.2018nsm,Selmke.2018c4lp}.
As the delay increases, the amplitude of this diffusive motion grows until a critical delay is reached and the particle begins to rotate around the target (see Supplementary Movies 1--3). We characterize this dynamics with the help of the angle $\theta$ between the direction of motion in Eq.~\ref{eq:target_cohesion} and the negative radial direction $-\mathbf{r}(t)$ (see Fig.~\ref{fig:figure2}A). Note that $\sin(\theta)$ corresponds to the contribution of a single active particle $i$ to the rotational order parameter commonly used to describe collective rotational motion, i.e., $o_{R,i}=(\hat{{\bf r}}_i\times \hat {\bf u}_i)\cdot {\bf e}_z=\sin(\theta_i)$ \cite{Tunstroem.2013,Baeuerle.20204d8}.
Fig.~\ref{fig:figure2}A shows the experimental trajectories of this propulsion angle $\theta$ for a single active particle with $v_0=2.16\,\ms$ and three different delays. For short delays, $\theta$ fluctuates with a small amplitude around zero (Fig.~\ref{fig:figure2}A top). The fluctuations increase with the delay and lead to a flat-top probability density of the propulsion angle for $\delta t \approx 0.87\, {\rm s}$ (Fig.~\ref{fig:figure2}A middle). At larger delays ($\delta t=1.14\, {\rm s}$), the propulsion angle fluctuates around a stable nonzero value which intermittently changes between a clockwise and counter-clockwise rotation sense of rotation (Fig.~\ref{fig:figure2}A bottom). In this case, the probability density $p(\theta)$ becomes bi-modal (Fig.~\ref{fig:figure2}C). The periods of the swimmer staying in either stably rotating state increase in duration when the delay is further increased. At $\delta t=1.4\, \si{\second}$, the propulsion angle fluctuates around $\pm 80 \si{\degree}$. Under these conditions, the cohesion of the particle to the target becomes weak as the particle velocity is almost tangential to the target particle circumference. As a result, the distance $R$ of the particle to the target starts to fluctuate more strongly, as shown in the position histograms in Fig.~\ref{fig:figure2}B.

\begin{figure*}[h!]
    \centering
    \includegraphics[width=0.65\columnwidth]{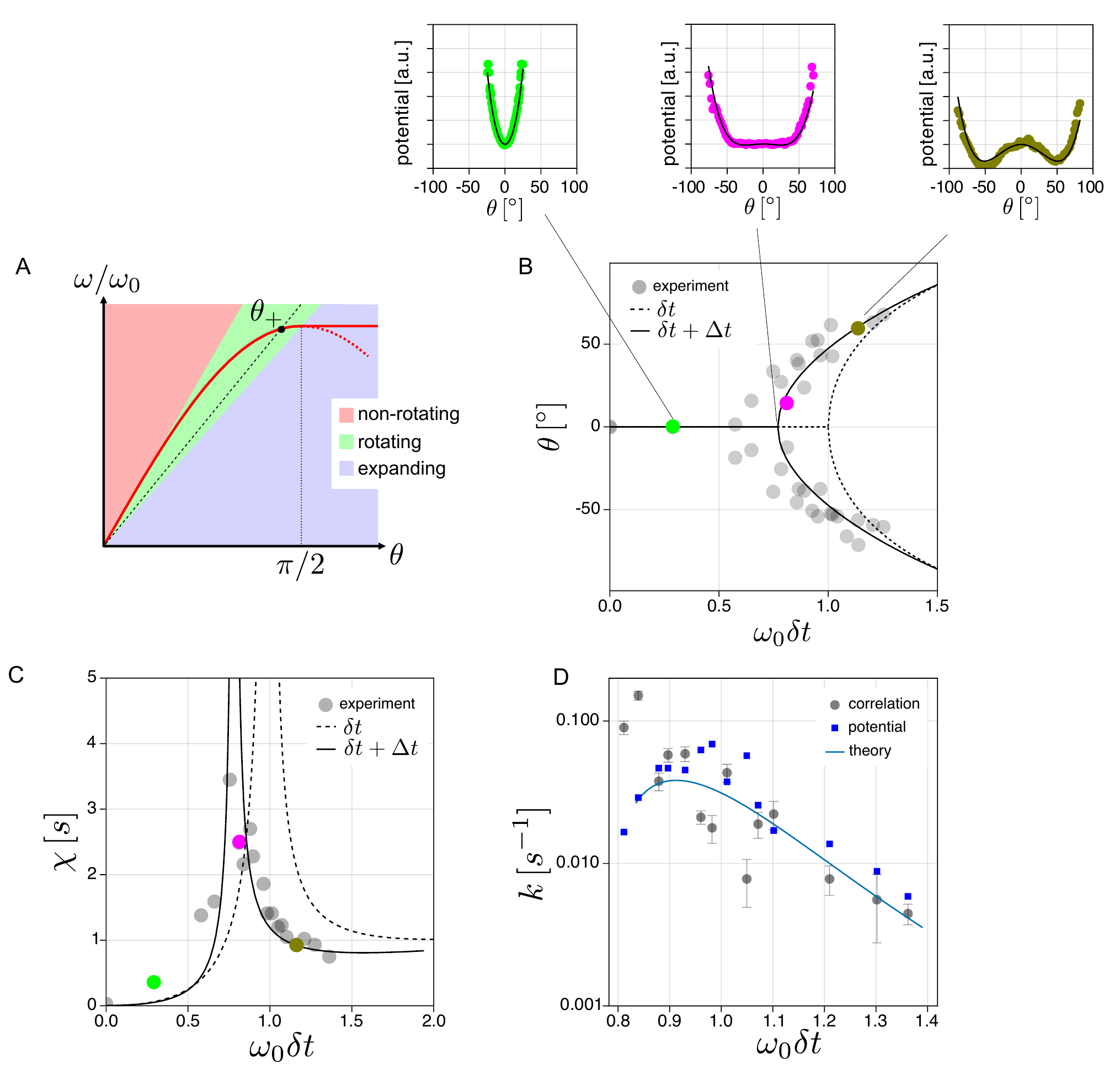}
    \caption{{\bf\small Disorder--order transition of a single active particle:} {\bf A} Graphical construction of the condition~\eqref{eq:simple} for a transition from a non-rotating state (red shaded region) to a rotating state (green shaded region) of a single active particle. The red line ($\sin \theta$) and the black dashed line with slope $1/(\omega_0 \delta t)$ intersect at several $\theta$. The solution $\theta=\theta_+$ in the green region and the corresponding solution $\theta_-$ in the 3rd quadrant (not shown) correspond to the co- and counter-clockwise rotating states. {\bf B} Experimentally obtained propulsion angles (maxima of the histograms in Fig.~\ref{fig:figure2}C) as a function of $\omega_0\delta t$, exhibiting bifurcation at $\omega_0\delta t\approx 0.76$. The solid line corresponds to the analytical prediction of the theoretical model~\eqref{eq:roots}. The dashed line shows the solution of the refined theoretical model, which includes the instrumental delay $\Delta t$ = 64\,ms of our setup in addition to the programmed delay $\delta t$. The colored dots denote the delays used in Fig.~\ref{fig:figure2} and the corresponding potentials of mean force, determined from the propulsion angle histograms in Fig.~\ref{fig:figure2}C, together with a fit of a refined analytical model including the instrumental delay $\Delta t$ (see SI). The only free parameter for fitting is the effective temperature of the system. {\bf C} Susceptibility $\chi$ of a single active particle as determined experimentally from the autocorrelation of the propulsion angle fluctuations (Eq.~\eqref{eq:autocorr}, data points). The solid lines correspond to the refined version of the theoretical prediction (Eq.~\eqref{eq:susceptibility}), including the instrumental delay $\Delta t$ (see SI for details). The colored dots have the same meaning as in the panel {\bf B}. 
    {\bf D} Transition rates between the two rotating states obtained from the experiments (circles) plotted with the predictions from Kramers' theory, Eq.~\eqref{eq:rate}, with $D_\theta=0.05$\,\si{\second}$^{-1}$ (solid line) and $D_\theta$ fitted to the probability distribution $p(\theta)$ separately for each value $\omega_0\delta t$ (squares).
    }
    \label{fig:figure3}
\end{figure*}

The non-zero propulsion angle is the result of an angular displacement during the period $[t-\delta t,t]$ of the active particle, and we can write it as
\begin{equation}
    \label{eq:deftheta}
    \theta(t) = \int^t_{t-\delta t} \omega(t') \dif t' = \phi(t) - \phi(t-\delta t) = \angle (\hat{\mathbf{u}}(t),-\mathbf{r}(t)).
\end{equation}
Here, $\phi(t)$ is the polar angle of the active particle in polar coordinates centered in the target particle and $\omega(t) = \dot \phi(t)$ is the corresponding instantaneous angular velocity (Fig.~\ref{fig:figure2}A). The observed dynamics can be understood by considering the active and target particles in physical contact. Their distance is then constrained to be the sum of their radii ($R=2a$). In this case, the active particle moves along the circumference of the target particle with an angular velocity $\omega(t)$ according to $\omega(t) =\omega_0\sin( \theta(t))$, where $\omega_0= v_0/R$ and the angle $\theta(t)$ is defined in Eq.~\eqref{eq:deftheta}. The term $\omega_0 \sin(\theta(t))$ originates from projecting the propulsion direction in Eq.~\eqref{eq:target_cohesion} parallel to the surface of the target particle at the point of contact. As sketched in Fig.~\ref{fig:figure3}A, assuming a constant angular velocity $\omega$ with $\theta = \omega\delta t$, the solutions to the equation for $\theta$ are given by the intersections of a sine function and a linear function,
\begin{equation}\label{eq:simple}
    (\omega_0 \delta t)^{-1}\theta=\sin(\theta).
\end{equation} 
For $\omega_0\delta t <1$, there is a single intersection at $\theta$ = 0, indicating a non-rotating state. For $1<\omega_0\delta t<\pi/2$, one non-rotating and two rotating solutions with opposite rotation senses arise. These rotations are stable under perturbations (see SI), while the non-rotating solution is unstable. For $\omega_0\delta t >\pi/2$, the rotating solutions correspond to $|\theta|>\pi/2$, and the radial component of propulsion becomes positive (repulsive), driving the active particle away from the target particle. As a result, the radius of the orbit, $R$, increases until a new stable orbit with $R = 2 v_0 \delta t/\pi>2a$ and $|\theta|=\pi/2$ is reached. For infinitely small particles ($a\to 0$), the distance of the swimmer to the target position can thus, in principle, vanish ($R\to 0$), and the rotation can occur at infinitely short programmed delays ($\delta t\to 0$). The retarded attraction hence always leads to rotation, and the observed transition from the non-rotating to the stable rotating state is caused by the minimum distance to the target. In the experiment, the minimum distance is determined not only by the particle radius $a$ but also by the Brownian motion of the active particle within the instrumental delay of the feedback loop, and it is nonzero even for $a>0$ \cite{Khadka.2018}.

Adding Brownian fluctuations  to the deterministic motion of the active particle described by Eq.~\ref{eq:simple}, results in the non-linear delayed stochastic differential equation 
$\dot{\phi}(t) = \omega_0 \sin\left(\phi(t)-\phi(t-\tau)\right) + \sqrt{2D_0/R^2} \, \eta(t)$, where $D_0 \approx 0.0642$\,{\si{\micro\metre^\, \second^{-1}}} denotes the translational diffusion coefficient of the active particle and $\eta(t)$ white noise. This equation is very difficult to solve. However, as detailed in the supplementary information, after several approximations it yields the simple overdamped Langevin equation
\begin{equation}
\dot{\theta} =\frac{1}{3 \delta t}\left[\theta_{\pm}^2 - \theta^2\right]\theta 
+ \sqrt{2D_\theta}\eta
\label{eq:domega_approx_overdamped}
\end{equation}
with  
\begin{equation}
\theta_{\pm} = \pm \sqrt{\frac{6}{\theta_0}(\theta_0 - 1)}
\label{eq:roots}
\end{equation} 
and $\theta_0 \equiv \omega_0 \delta t$. The noise term in Eq.~\eqref{eq:domega_approx_overdamped} describes the angular Brownian motion of the active particle around the target with an effective diffusion coefficient $D_\theta$, which is a single free parameter in our theory. Eq.~\eqref{eq:domega_approx_overdamped} yields the stationary solutions $0$ and $\theta_{\pm}$, consistent with our previous simplified model. All the three solutions coincide when $\theta =\theta_0 = \omega_0\delta t = 1$, implying a transition from a non-rotating to two rotating states. The data points in Fig.~\ref{fig:figure3}B display the experimentally obtained maxima of the histograms $p(\theta)$ of the propulsion angle (see Fig.~\ref{fig:figure2}C) as a function of $\omega_0\delta t$. While the transition to the two rotating states is expected at $\omega_0\delta t$ = 1, the transition points in experiments are located at lower values due to an additional instrumental delay $\Delta t$ in the feedback loop of the experimental setup. This instrumental delay between the most recent exposure to the camera and the laser positioning affects the motion direction \cite{Fraenzl.2021,Muinos-Landin.2021}, causing an earlier onset of the transition to a stable rotation. The dashed line in Fig.~\ref{fig:figure3}B shows the theoretical prediction, which includes the instrumental delay $\Delta t$ and the programmed delay $\delta t$, as detailed in the SI (Eq.~(17)).
The Langevin equation~\eqref{eq:domega_approx_overdamped} can be interpreted as a dynamical equation for the position $\theta$ of an overdamped Brownian particle with friction and diffusion coefficients $\gamma$ and $D_\theta$ in a quartic potential (see derivation in the SI), 
\begin{equation}
U(\theta)=\frac{\gamma}{\delta t}\left[ \left(\frac{1}{\theta_0}- 1\right )\theta^2 +\frac{1}{12}\theta^4 \right],
\label{eq:potential}
\end{equation}
which corresponds to the generic case of a supercritical pitchfork bifurcation \cite{Strogatz.1994}. The potential can be extracted from the experimental data (Fig.~\ref{fig:figure3}B) by fitting the histogram $p(\theta)$ with the Boltzmann distribution $\exp(- U(\theta)/ \gamma D_\theta)/Z$ including an effective temperature $\gamma D_\theta/k_{\rm B}$ and normalization factor $Z$. The effective temperature thus represents a scaling factor that brings the measured potential of mean force $- \gamma D_\theta \log p(\theta)$ to the theoretical prediction in Eq.~\eqref{eq:potential}.

The self-generated quartic potential in Eq.~\eqref{eq:potential} emerging from the delayed response finds its mathematical analogue in the free energy function in Landau's theory of second-order phase transitions~\cite{Goldenfeld.2018}, with the control parameter $\theta_0$. This correspondence suggests that $\mathcal{T} =  (\omega_0\delta t)^{-1}=\theta_0^{-1}$ is the equivalent of a temperature with a critical point $\mathcal{T}_C$ = 1. Both the activity $\omega_0$ and the delay $\delta t$ in the product can lead to the transition. Hence, the active particle velocity and delay are inversely coupled, and large particle velocities need only short delays causing an ordered rotating state. Yet, our system does not correspond to a thermodynamic phase, where the transitions are of collective nature. The bifurcation and the potential are caused by the delayed response of the single active particle to an external signal including the steric repulsion by the target particle.
A stability analysis of Eq.~\eqref{eq:domega_approx_overdamped} (see SI) yields the susceptibility 
\begin{equation}\label{eq:susceptibility}
    \chi\left (\frac{1}{\theta_0} \right) =
    \begin{cases} 
    \frac{\delta t}{2\left (\frac{1}{\theta_0}-1\right )} & \text{if $\frac{1}{\theta_0}>1$}\\ 
    -\frac{\delta t}{4\left (\frac{1}{\theta_0}-1\right )} &  \text{if $\frac{1}{\theta_0}<1$} 
    \end{cases}\;,
\end{equation}
as familiar from Landau theory~\cite{Goldenfeld.2018} (for $\gamma = 1$\,s). The susceptibility corresponds to the relaxation time of the system after a perturbation from a stable state (see SI). It is determined from the experiments by measuring the autocorrelation function
\begin{equation}\label{eq:autocorr}
    C(\tau)=\frac{\langle \delta \theta(t+\tau) \delta \theta(t)\rangle_t}{\langle\delta \theta(t)^2 \rangle_t}
\end{equation} 
of fluctuations of the propulsion angle $\delta \theta(t)=\theta(t)-\langle\theta(t)\rangle$.

Fig.~\ref{fig:figure3}C shows the decay time $t_R$ obtained from the autocorrelation function as $C(t_R)$ = 1/e (symbols) together with susceptibilities predicted theoretically without (Eq.~\eqref{eq:susceptibility}, dashed line) and with (see SI, solid line) instrumental delay $\Delta t$. Close to the transition point ($\omega_0 \delta t \approx 1$), the theory suggests a slowing down of the relaxation due to an increasingly flat potential (e.g., the potential plot in the middle of Fig.~\ref{fig:figure3}B). The experimental results nicely confirm this prediction without any fitting parameter. In addition to calculating the susceptibility from the autocorrelation function, we also performed relaxation measurements after a step-like perturbation of the propulsion angle $\theta$. The results of the measurements agree well with the theoretical predictions, as shown in Sec. 4 of the SI, together with details on the evaluation of the susceptibilities.

While both stable rotation directions can be inferred from a purely deterministic model excluding Brownian motion, the observed spontaneous reversal of the rotation direction is driven by fluctuations in the propulsion angle and thus by the noise in the system. The changes in the rotation direction correspond to transitions between the minima $\pm\theta_{\pm}$ of the self-generated potential, Eq.~\eqref{eq:potential}. We may thus apply Kramers' theory to estimate the corresponding transition rate as
\begin{equation}
k =  \frac{\sqrt{2}}{\pi}\frac{|\theta_0-1|}{\theta_0 \delta t}\exp\left[-\frac{3}{\gamma D_\theta}\frac{(\theta_0 - 1)^2}{\theta_0^2}\right].
\label{eq:rate}
\end{equation}
The effective temperature $\gamma D_\theta/k_{\rm B}$ driving the fluctuations in the potential is obtained from the previously mentioned scaling of the theoretical prediction to the measured potential of mean force. Fig.~\ref{fig:figure3}D displays the results of experiments measuring the transition rate from mean residence times of $\theta$ in the two potential wells. The rate is compared to the prediction from the Kramers theory, Eq.~\eqref{eq:rate}, and shows a good agreement.

\subsection*{Multiple Particles}

As demonstrated in the previous section, the rotation observed in our experiments results from a spontaneous symmetry breaking in the dynamics of a single active agent under a retarded self-propulsion to a target, different from standard models of rotational dynamics in overdamped systems, which assume mutual or "social" interactions among the agents \cite{Vollmer.2006,Costanzo.2018,Berdahl.2018i5,Delcourt.2016}. Hence, when more active particles are added to the system, each of them strives to exhibit the same rotation and bifurcation as the single swimmer. However, steric, hydrodynamic, and thermophoretic interactions among the particles synchronize and stabilize their motion so that the system exhibits collective behavior.

\begin{figure*}[h!]
    \centering
    \includegraphics[width=0.8\columnwidth]{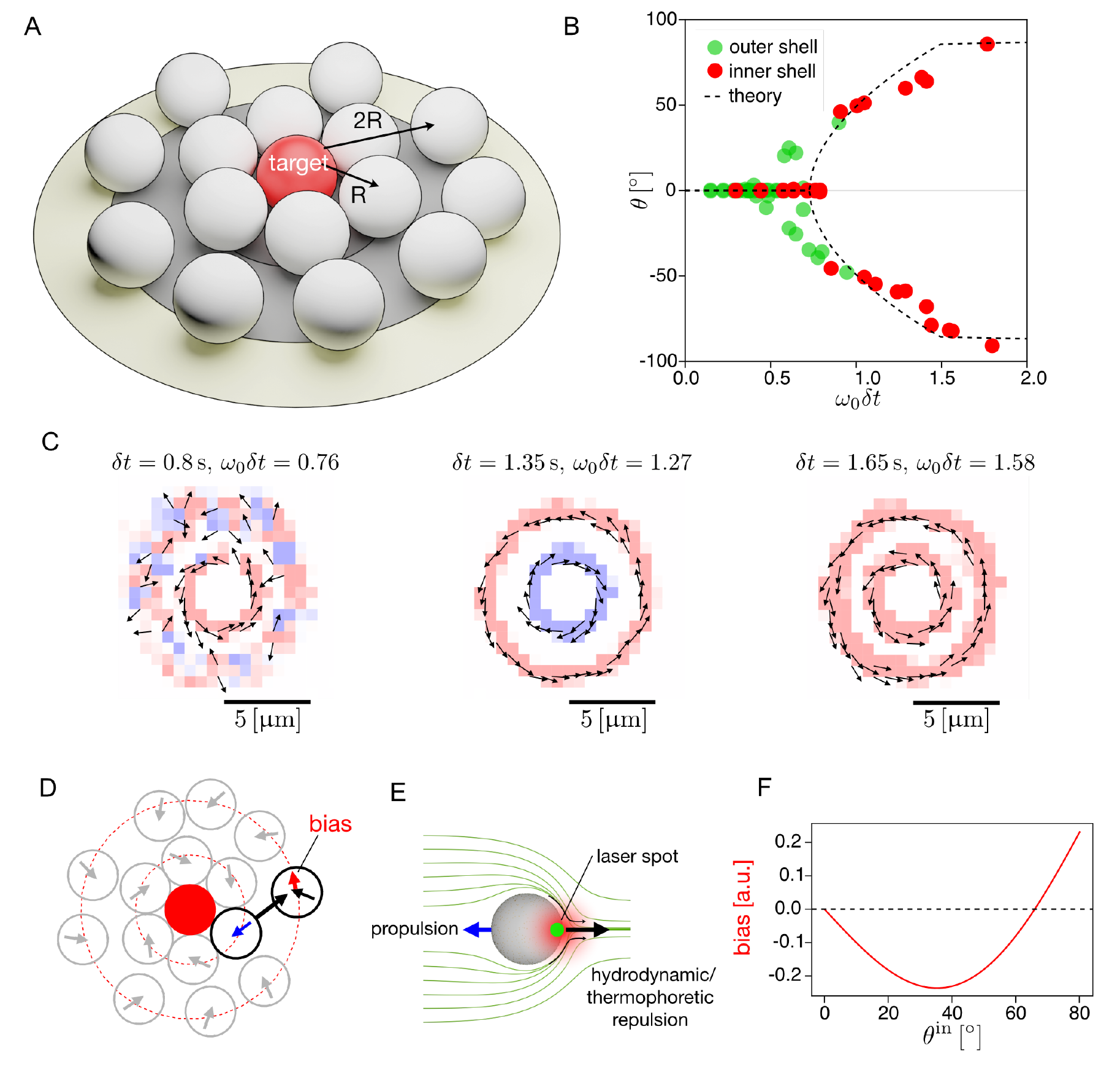}
    \caption{{\bf Collective rotation of 15 particles attracted to a single target particle:} {\bf A} Sketch of the shell structure and radii. {\bf B} Bifurcation of the most probable propulsion angle as a function of $\omega_0 \delta t$ for the active particles with the speed of $v_0=2.06\, \ms$ (calibrator speed).  The red dots are obtained from the inner shell particles at a typical distance of $R^\mathrm{in}=2.18\, {\rm \mum}$, while the green dots denote the outer shell particles at $R^\mathrm{out}=4.47\, {\rm \mum}$. The dashed line corresponds to the theoretical single particle prediction including the instrumental delay $\Delta t$ = 70\,ms. 
    {\bf C}  Average velocity field of active particles at
    $\delta t$ = 0.81\,s when the spontaneous rotation of the inner shell is constantly disrupted by the non-rotating outer shell, $\delta t$ = 1.35\,s when the two shells are rotating in opposite directions, and at $\delta t$ = 1.65\,s when both shells are rotating in the same direction. The arrows and colors denote the average direction of motion. 
    {\bf D} Snapshot of the active particles and their propulsion directions corresponding to 
    {\bf C}, $\delta t$ = 1.35\,s. The repulsion induced by the flow and temperature fields of the inner shell causes a bias for the outer shell rotation. 
    {\bf E} Sketch of the flow and temperature fields around an active particle induced by the laser (green dot), and the resulting repulsion. 
    {\bf F} Sketch of the presumable magnitude of the bias caused by the temperature and flow fields on the rotation of the outer shell as a function of the propulsion angle of the inner shell particles $\theta^\mathrm{in}$ (see SI).
    \label{fig:figure4}}
\end{figure*}

Fig.~\ref{fig:figure4} summarizes the key results obtained for an ensemble of 15 active particles attracted to the target particle with the same delay.
For the considered range of time delays, the active particles form two tightly packed shells around the target particle (Fig.~\ref{fig:figure4}A).
The typical distance of the inner shell particles to the target is about half that of the outer shell, $R^\mathrm{out} \approx 2R^\mathrm{in} = 4a$. So the inner shell is expected to enter the rotational phase with a correspondingly shorter delay than the outer shell. However, the interparticle interactions in the compact cluster strongly correlate the particle motion, which quantitatively changes the results obtained for a single particle. 
Compared to the theoretical prediction, $\omega_0 \delta t$ = 0.73, we observe that for $v_0$ = 2.06\,$\ms$ the transition to the rotational phase of the inner shell is postponed to $\omega^{\mathrm{in}}_0 \delta t \equiv v_0\delta t/ R^\mathrm{in}\approx$ 0.83, corresponding to $\delta t$ = 0.9\,s (see the rightmost red data point lying on the horizontal axis in Fig.~\ref{fig:figure4}B). Close to the transition, the inner shell exhibits alternating periods of rotation and non-rotation, whereas the non-rotating outer shell compresses the inner shell due to its inwards pointing propulsion direction (Fig.~\ref{fig:figure4}C, left).

Fig.~\ref{fig:figure4}C displays the velocity fields of the particles averaged over their trajectories with three different delays. 
The rotation of the outer shell is observed at $\omega^{\mathrm{out}}_0\delta t \equiv v_0\delta t/ R^\mathrm{out} \approx$ 0.41, corresponding to the same delay $\delta t$ = 0.9\,s when the inner shell starts to rotate (see Fig.~\ref{fig:figure4}B and Supplementary Movies 4--6). For delays slightly above the transition, $0.9<\delta t$ $<$ 1.41\,s, the two shells rotate in opposite directions, as shown in the middle plot of Fig.~\ref{fig:figure4}C. The simultaneous transition and the counter-rotation of the two shells suggest that the inner shell particles generate forces in the opposite direction to their propulsion, repelling the outer shell particles and causing a bias on their rotation, as depicted in Figs.~\ref{fig:figure4}D--F. 

The bias is presumably caused by the directional hydrodynamic and thermophoretic interactions. The surface temperature gradient across the particle creates a thermo-osmotic surface flow that propels the particle \cite{Bregulla.201633m}. If the particle motion is (partly) halted by an external force such as the steric force applied by the immobilized target particle, the particle acts as a pump creating a hydrodynamic outflow at the hot side of the swimmer (Fig.~\ref{fig:figure4}D and SI). Similarly, thermophoretic interactions arise from temperature gradients across the surface of an active particle caused by its neighboring particles \cite{Auschra.2021pnk}. They are commonly repulsive as found, e.g., for Janus particles in external temperature gradients \cite{Auschra.2021pnk}. We have carried out finite element simulations of the flow field around a mobile and an immobile self-propelling swimmer (see SI). The overall near field hydrodynamic interactions are nevertheless complex due to the presence of the other particles and the substrate surface \cite{Popescu.2017ut,Spagnolie.2012,Lintuvuori.2017}, and further depend on the propulsion angle $\theta$.
The inner shell propulsion angle increases with the delay, which results in changing direction and magnitude of the inter-shell interaction and thus the outer shell rotation bias, which presumably varies as sketched in Fig.~\ref{fig:figure4}F (see SI). For $\delta t \geq 1.41$\,s, the two shells predominantly rotate in the same direction, as shown in Fig.~\ref{fig:figure4}C, right. The transition from counter- to co-rotation shells corresponds to the sign flip of the bias, occurring at $\theta^\mathrm{in}\approx$ 67$^\circ$. 
While we currently cannot separate thermophoretic and hydrodynamic effects in the experiment, possible hydrodynamic interactions yield an interesting perspective as they are supposed to influence also the collective behavior of fish or birds. 
At longer retardation, $\theta^\mathrm{in}$ tends to reach 90$^\circ$, and thus the inner shell tries to expand against the compressing outer shell. These competing tendencies lead to particle exchange between the two shells. 

\section*{Discussion}

We have demonstrated above that the motion of an active particle  induced by the delayed attraction to a target point can spontaneously undergo a transition from a diffuse isotropic state to a dynamical chiral state, upon increasing the activity and/or the delay time. The transition is well described by a pitchfork bifurcation accompanied by a characteristic critical slowing down of the response \cite{Strogatz.1994}. The single-particle dynamics thus already exhibits non-trivial features more commonly associated with (mean-field) phase transitions in strongly interacting passive many-body systems~\cite{Fletcher.1997}. This can be explained by noting that the deterministic part, 
$
    \dot{\phi}(t) =\omega_0\sin\left (\phi(t)-\phi(t-\delta t)\right)
$,
of our stochastic delay differential equation can also be understood as the dynamical equation for a single Kuramoto phase oscillator \cite{Kuramoto.1975,OKeeffe.2017}, with vanishing eigenfrequency and coupling strength $\omega_0$, which is trying to synchronize with its own past state. In the chiral state, the particle orbits around the central obstacle. This orbiting motion is stable against noise, but its chirality is only transiently maintained. This should be contrasted to the chiral states resulting from non-reciprocal coupling in the Kuramoto model as discussed by Vitelli et al. \cite{Fruchart.2021}, which require many-body cooperativity. Despite the similarity of the two models, we are clearly observing a single-particle effect, instead. The lack of many-body cooperativity is compensated for by the infinite number of relaxation modes encoded in the time-delayed equation of motion \cite{Loos2019, Geiss.2019}. 

The nonlinear dynamics of our experimental system can be described by an approximate analytical model resulting in an emergent self-generated quartic potential. While such potentials are frequently found in descriptions of phase transitions and collective effects in active particle ensembles, following various behavioral rules \cite{Baeuerle.20204d8,Loeffler.2021}, this is not the case, here. Due to the activity and the (programmed) delay, it already occurs for a single active particle aiming at a spatially fixed target. In a whole swarm of particles  that are all attracted to a common target, which might be its own perceived center of mass, the single-particle bifurcation is preserved. Inter-particle collisions merely synchronize and stabilize the rotational states of the individual particles. Upon close contact, hydrodynamic and thermophoretic interactions become important and help the  swimmers to self-organize into co- and counter-rotating orbits. In  biological motile ensembles, from bacteria to fish, similar hydrodynamic mechanisms may be at work, although precise details and scales may differ widely \cite{Drescher.2011,Spagnolie.2012,Lauder.2002,Verma.2018g4d}. The corresponding many-body effects can be subtle and may elude coarse grained simulations and theories. This underscores the importance of well controlled experimental model systems that may act as ``hybrid simulations'', combining computer-controlled active particles with real-world environments. 

While time delays are an unavoidable outcome of coarse-graining microscopic descriptions of the feedback processes in natural systems, they are often neglected in low-dimensional models of active particle collective effects \cite{Vicsek.1995,Pearce.2014}. In this respect, our model system provides a new perspective, as it takes the unavoidable systematic delays in the dynamics seriously and explores their generic effects. It thereby provides a microscopic underpinning for empirical vision-cone models that have been found to produce rotational dynamics in overdamped flocks or swarms \cite{Baeuerle.20204d8,Bregulla.2014}.
In overdamped systems, retardation thus plays a similar role as added inertia. 
Both effects lead to persistence in the particle dynamics, but the effects of the time delay may be much richer~\cite{Mijalkov.2016,Geiss.2019,Holubec.2021}.

While we considered only a positive delay, i.e., synchronization with the past, sophisticated biological organisms also have predictive capabilities at their disposal that allow them to extrapolate the current state into the future \cite{Palmer.2015}. These can to a first approximation be incorporated in the form of a negative time delay. 
The inclusion of positive and negative delays may therefore also provide a new perspective on claims that collision and alignment interactions, as well as other complex rules, are the sources of the emerging complex adaptive responses observed in living many-body systems. 

\section*{Methods}
\subsection{Sample Preparation} Samples were prepared using two glass coverslips (20\,mm\,$\times$\,20\,mm, 24\,mm\,$\times$\,24\,mm) to confine a thin liquid layer (3\,$\mum$ thickness) in between. The edges of one coverslip are sealed with a thin layer of PDMS (polydimethylsiloxane) to prevent leakage and evaporation.
The liquid film used in the sample is composed of 2.19\,$\mum$ diameter gold-coated melamine formaldehyde (MF) particles (microParticles GmbH) dispersed in 0.1\% Pluronic F-127 solution. The latter prevents the cohesion of the particles and adsorption to the cover slide surface. The surface of the MF particles is uniformly scattered with gold nano-particles of about 8\,nm diameter with a total surface coverage of about 10\% (Fig.~S3A).
SiO$_2$ particles (2.96\,$\mum$ in diameter, microParticles GmbH) are added into the solution to keep the thickness of the liquid layer at about 3\,$\mum$. 
Finally, 0.3~$\mathrm{\mu l}$ of the mixed particle suspension is pipetted on one of the coverslips, and the other is put on top.

\subsection{Experimental Setup} The experimental setup (see SI) consists of an inverted microscope (Olympus, IX71) with a mounted piezo translation stage (Physik Instrumente, P-733.3). The sample is illuminated with an oil-immersion darkfield condenser (Olympus, U-DCW, NA 1.2--1.4) and a white-light LED (Thorlabs, SOLIS-3C). The scattered light is imaged by an objective lens (Olympus, UPlanApo $\times$100/1.35, Oil, Iris, NA 0.5--1.35) and a tube lens (250\,mm) to an EMCCD (electron-multiplying charge-coupled device) camera (Andor, iXon DV885LC). The variable numerical aperture of the objective was set to a value below the minimum aperture of the darkfield condenser. 

The microparticles are heated by a focused, continuous-wave laser at a wavelength of 532\,nm (CNI, MGL-III-532). The beam diameter is increased by a beam expander and sent to an acousto-optic deflector (AA Opto-Electronic, DTSXY-400-532) and a lens system to steer the laser focus in the sample plane. The deflected beam is directed towards the sample by a dichroic beam splitter (D, Omega Optical, 560DRLP) and focused by an oil-immersion objective (Olympus, UPlanApo $\times$100/1.35, Oil, Iris, NA 0.5--1.35) to the sample plane ($w_0 \approx 0.8\,\mum$ beam waist in the sample plane). A notch filter (Thorlabs, NF533-17) is used to block any remaining back reflections of the laser from the detection path. 
The acousto-optic deflector (AOD), as well as the piezo stage, are driven by an AD/DA (analog-digital/digital-analog) converter (Jäger Messtechnik, ADwin-Gold II). A LabVIEW program running on a desktop PC (Intel Core i7 2600 4 $\times$ 3.40~ GHz CPU) is used to record and process the images as well as to control the AOD feedback via the AD/DA converter.

\section*{Acknowledgement}
 We acknowledge the support of the COVID-19 pandemic. We acknowledge funding through
a DFG-GACR cooperation by the Deutsche Forschungsgemeinschaft (DFG Project No 432421051) and by
the Czech Science Foundation (GACR Project No 20-02955J). VH was supported by the Humboldt Foundation. We thank Andrea Kramer for proof-reading the manuscript.

\section*{Competing interests}
The authors have no competing interests.

\section*{Author Contributions}
XW and FC conceived the experiment. XW carried out the experiment. XW, FC, PC, VH analyzed the data. PC and VH developed the theory. XW, PC and FC carried out simulations. All authors discussed the results and wrote the manuscript.

\section*{Data availability}
All data in support of this work is available in the manuscript or the supplementary materials. Further data and materials are available from the corresponding author upon request.

\bibliography{references.bib}

\end{document}